\documentclass[12pt]{article}
\usepackage{epsfig, amssymb, graphicx, amsmath} 
\usepackage{amsthm}
\usepackage[round-mode=places, round-integer-to-decimal, round-precision=2,
    table-format = 1.2, 
    table-number-alignment=center,
    round-integer-to-decimal,
    output-decimal-marker={,}
    ]{siunitx} 
\usepackage{booktabs}
\usepackage{enumitem}
\usepackage{eurosym}	
\usepackage{wrapfig}	
\usepackage{float}		
\usepackage{subfig}	
\usepackage{footnote}
\usepackage{thmtools}
\usepackage{bbm}		
\usepackage[title]{appendix}
\usepackage{cleveref}
\usepackage[authoryear]{natbib}
\bibpunct{(}{)}{,}{a}{}{,}

\theoremstyle{definition}

\theoremstyle{remark}

\numberwithin{theorem}{section}
\numberwithin{proposition}{section}
\numberwithin{lemma}{section}
\numberwithin{corollary}{section}
\numberwithin{definition}{section}
\numberwithin{remark}{section}
\numberwithin{example}{section}

\newcommand{\be}{\begin{equation}}
\newcommand{\en}{\end{equation}}
\newcommand{\ben}{\begin{equation*}}
\newcommand{\enn}{\end{equation*}}
\newcommand{\bea}{\begin{eqnarray}}
\newcommand{\ena}{\end{eqnarray}}

\begin{document}
 
\newlength\tindent
\setlength{\tindent}{\parindent}
\setlength{\parindent}{0pt}
\renewcommand{\indent}{\hspace*{\tindent}}

\begin{savenotes}
\title{
\bf{ 
 Synthetic forwards and cost of funding \\
in the equity derivative market 
}}
\author{
Michele Azzone$^\ddagger$ \& 
Roberto Baviera$^\ddagger$ 
}

\maketitle

\vspace*{0.11truein}
\begin{tabular}{ll}
$(\ddagger)$ &  Politecnico di Milano, Department of Mathematics, 32 p.zza L. da Vinci, Milano \\
\end{tabular}
\end{savenotes}

\vspace*{0.11truein}
\begin{abstract}
This study introduces a new technique 
to recover the implicit discount factor in the derivative market 
using only European put and call prices:
this discount is grounded in actual transactions in active markets.
Moreover, this study identifies the 
 implied cost of funding, over OIS, 
of major market players. 

\noindent 
Does a liquid equity market allow arbitrage? The key idea is that  the (unique) 
forward contract -built using the put-call parity relation- contains information about the market discount factor: by no-arbitrage conditions we identify the implicit interest rate such that the forward contract value does not depend on the strike.

\noindent  
The procedure is applied to options on S\&P 500 and EURO STOXX 50 indices. 
There is statistical evidence that, in the EURO STOXX 50 market, the implicit interest rate curve coincides 
with the EUR OIS one,
while, in the S\&P 500 market, a cost of funding of, on average, 34 basis points is added on top of the USD OIS curve.
\end{abstract}

\vspace*{0.11truein}
{\bf Keywords}: 
Forward price, put-call parity, implied interest rate, cost of funding, synthetic forward
\vspace*{0.11truein}

{\bf JEL Classification}: 
E43, G12, G13. 

\vspace*{1cm}
\begin{flushleft}
	Cite as: Azzone, A. \& Baviera, R. (2021). Synthetic forwards and cost of funding in the equity derivative market. \textit{Finance Research Letters}, \textit{41}. \\
\end{flushleft}

\vspace{2cm}
\begin{flushleft}
{\bf Address for correspondence:}\\
Roberto Baviera\\
Department of Mathematics \\
Politecnico di Milano\\
32 p.zza Leonardo da Vinci \\ 
I-20133 Milano, Italy \\
Tel. +39-02-2399 4630\\
Fax. +39-02-2399 4621\\
roberto.baviera@polimi.it
\end{flushleft}

\newpage

\begin{center}
\Large\bfseries 
 Synthetic forwards and cost of funding \\
in the equity derivative market
\end{center}

\vspace*{0.21truein}

\section{Introduction}

The term structure of interest rates is a crucial input in the derivative market. It is used for determining the discount rate for expected payoffs in a given currency. 

The main research question we consider in this letter is: when dealing with liquid exchange-traded derivatives, which is the interest rate term structure used by market makers?

In general, interest rates used in derivative pricing are not “risk-free” because contingent claim evaluation should depend on the risks of the investment and in particular, 
on the funding risk and 
on the risk of default of one of the two counterparties in the derivative contract.\footnote{For a dealer, 
the expected loss due to a possible default by the counterparty is related to the credit value adjustment 
and the expected gain due to a possible default by the dealer itself is referred to as the debt value adjustment; 
the funding risk is associated with the funding valuation adjustment. 
This pricing approach can be found in excellent textbooks \citep[see, e.g.][]{Gregory, BMP2013}.}

\bigskip

When dealing with exchange-traded derivatives the situation should be simpler: 
the presence of a clearinghouse
with margin calls allows neglecting the market participants' default risk.
Before the Great Financial Crisis of 2007, the answer was to consider the Libor curve as the discounting term structure \citep[see e.g.][p.14]{HullWhite2013_LiborOIS}. 
After the crisis, the difference between Libor rates with different tenors enlarged to several tenths of basis points \citep[see, e.g.][and references therein]{HenrardBook} making this answer less obvious; more recently the situation has become even more complicated, in particular after July 2017, when the Chief Executive of U.K. Financial Conduct Authority (FCA) 
in a famous speech \citep{Bailey} increased market expectation that Libor benchmarks will be discontinued within a few years \citep[see, e.g.][for a clear and exciting discussion on the Libor fallout from a quantitative perspective]{Henrard_Fallback2019}.

\bigskip

The Overnight Index Swap (OIS) curve has emerged as a possible candidate for the risk-free curve for derivative discounting in the aftermath of the crisis.
The OIS is a swap derived from the unsecured interbank overnight rate (OR), 
which is, for example, the EONIA rate for Euros\footnote{ Substituted by the Euro short-term rate (\euro{}STR) starting from the $2^nd$ of October 2019, i.e.
after the period of analysis considered in this letter.} and the Effective Federal Fund Rate (EFFR) for US dollars\footnote{USD OIS market is mainly based on this rate. The OIS swap trading volumes based on the Secured Overnight Financing Rate (SOFR) 
is negligible w.r.t. 
the total OIS volume at the time of writing.}. This OR can be considered a good proxy of a risk-free rate and it is the interest rate most commonly paid on margins.
Moreover, the OIS curve presents several advantages: it is a curve based on liquid swaps. The bootstrap of the discounting curve is as simple as the well-established pre-crisis methodology \citep[see, e.g.][]{Ron}. 

\bigskip

The approach of selecting the interest rate term structure from a practitioner perspective appears relatively clear.
We are particularly interested in a market maker
who operates in a given exchange-traded derivative market.
Often he considers the OIS curve for discounting, allowing for a spread that accounts for other risks or costs 
not included in the ``risk-free" rate.

We call this spread ``cost-of-funding'' because  it can be seen as the implicit additional cost in operating in this derivative market.
 We reformulate  our research question for this practitioner:
which is the cost of funding (if any) of operating in a liquid exchange-traded derivative market?

The answer to this question has both operational and management implications. 
On the one side, for his daily activity, the market maker should build and monitor an indicator on this spread, to use a discounting curve in line with other market participants; on the other side, this spread has relevant consequences on the management of a financial firm. 
If, within a financial institution,  
the cost of funding of a given business unit of market-making is larger than the market, it is rather difficult that this unit can be competitive in the derivative market where it operates.
Determine at which cost of funding each business unit should operate is a relevant management decision within a financial firm.
This letter introduces an elementary indicator that can monitor in real-time the funding cost and point out a possible stress in funding liquidity.

\bigskip

We consider all options on the EURO STOXX 50 and the S\&P 500, respectively the most liquid equity index in the Euro area and in the U.S.A. \citep[see e.g.][]{dash2009capturing,bai2019improving, vo2008volatility, niguez2016evaluating}. 

This study builds over the put-call parity of European options. 
The idea of using  put-call parity
to obtain the implied  interest rates dates back to \citet{brenner1986implied},
who consider underlyings that do not pay dividends.
\citet{frankfurter1991further} and \citet{naranjo2009implied} extended this methodology to options on an underlying that pays dividends. 
To apply their techniques it is necessary to know both the 
forward prices and the forward dividends. 
They infer future dividends from realized ones and discuss the differences between the discount factor observed in the market and the discount factor obtained from the LIBOR and Treasury curves.  \\

This study  presents an alternative approach that allows us to obtain the implicit interest rates 
 using only option prices and a no-arbitrage condition on an option portfolio known as {\it synthetic} forward.\footnote{ Synthetic forwards are perfectly synchronized with option prices. There is empirical evidence that, in some markets, they are more reliable than quoted futures \citep[see e.g.][]{muravyev2013there, hao2020price}.}
The implicit interest rates of the S\&P 500 and EURO STOXX 50 options markets are computed with a simple technique.
Together with OIS discounting term structure, this technique allows a market maker to build an elementary measure of the cost of funding that can be obtained instantly 
from option prices.

\bigskip

The rest of the letter is organized as follows.
Section \textbf{2} shows the methodology to find the implicit interest rates using only option prices and describes the dataset. 
Section \textbf{3} infers the S\&P 500 and the EURO STOXX 50 implicit discount factor and the corresponding cost of funding. 
Section \textbf{4} concludes.

\section{The methodology and the dataset}

This section shows how to obtain the discount factor from market data using only call and put prices. 
We present the dataset and discuss the data preprocessing techniques.\\

The absence of arbitrage condition allows us to write, at value date $t_0$ and at a fixed maturity $T$,  
the put-call parity for European options \citep[see, e.g.][Ch.8 p.174]{hull2003options}  w.r.t. the forward price $F$ and the strike price $K$
\begin{equation}
  C(K)-P(K) = \overline{ B}(t_0,T) (F-K) \;\;,
\label{eq:CallPutParity}
\end{equation}
where $C(K)$ and $ P(K)$ are respectively the European call and put option prices and $\overline{ B}(t_0,T)$ is the market discount factor between $t_0$ and $T$.

 Instead of considering a standard forward contract, a trader in this market can mimic this position
using call and put options with the same strike price and the same maturity to create a forward position: this position is called {\it synthetic} forward.
The synthetic forwards are frequently traded in the equity derivative markets:
they identify --for several maturities-- the most liquid forwards in the market.

A  synthetic forward ${\mathcal G}(K)$ with maturity $T$ is a portfolio that comprises of a long call and a short put at a given strike price $K$. Forward prices in $t_0$ with the same maturity $T$ are all equivalent whatever strike $K$ is considered
and, due to the no-arbitrage condition, they should have the same price.\footnote{We could build an arbitrage position on synthetic forwards with the same maturity and different strikes via the so-called \textit{box} strategy: i.e. a position composed by a long synthetic forward at a given strike and a short synthetic forward at a different one. For this strategy -that is equivalent to a long or short cash position- we can neglect margin (MVA) and capital (KVA) adjustment.}
The market implied discount factor $\overline{ B}(t_0,T)$ is the (unique) factor such that the 
forward price  
\begin{equation}
  F =\frac{ {\mathcal G} (K)}{\overline{ B}(t_0,T)}+K
\label{eq:Forward}
\end{equation}
does not depend on the strike $K$:
this is the main idea of the letter. 
This is a linear problem in $\overline{ B}$ and $F$. 
We discuss its solution in Section \textbf{3}. \\

We consider all quoted  S\&P 500 and EURO STOXX 50 option prices\footnote{ We consider the CBOE European options on the  S\&P 500 index (option prices are reported by the U.S.A. Options Price Reporting Authority) and the Eurex European options on the EURO STOXX 50 index. Eikon Reuters option chains are respectively \textit{0\#SPX*.U} and \textit{0\#STXE*.EX}.} observed at 3:00 
p.m. London Time  
 each business day from the $1^{st}$ of November 2018 to the $19^{th}$ of July 2019 excluding days from the $20^{th}$ of December 2018 to the $6^{th}$ of January 2019 and from the $13^{th}$ of April 2019 to the $2^{nd}$ of May 2019.
For both indices, the most liquid options expire on the third Friday of the first six months after the value date and then on  
March, June, September, and December in the front year and June and December in the next year. 
In the EURO STOXX 50 case also June and December contracts for the following years are 
available.\footnote{For each value date $t_0$ we observe 10 to 13 liquid synthetic forward maturities in the S\&P 500 market and 18 to 19 contracts' maturities in the EURO STOXX 50.}
In Table \ref{tab:stat descr} we provide the  descriptive statistics of some relevant quantities in the options' dataset. We report the number of strike for each maturity, the straddle position, $C(K)+ P(K)$, and the synthetic forward plus the strike, ${\mathcal G}(K)+K$.
\begin{center}
\begin{tabular} {ccccccc}
			\toprule
		\textbf{	market} & \textbf{quantity} &\textbf{mean}&\textbf{median} &\textbf{std }&\textbf{$\mathbf{q_{0.05}}$}&\textbf{$\mathbf{q_{0.95}}$}\\ 
			\toprule
						{	S\&P 500}&\#Strikes &  131 & 95&67 &77 &250 \\
								{	S\&P 500} &C+P &  $585.70$ & 488.00& $396.56$ &130.02 &1389.42 \\
		{	S\&P 500}&${\mathcal G}$+K & 2794.00&2796.75 &$118.61$& 2583.45& 2991.30\\

			\hline
						{	EURO STOXX 50}&\#Strikes  &44&46 & 19 &16 &73\\
									{	EURO STOXX 50}&C+P& $584.59$ & 533.55&$   331.27$ &172.90 &1226.35 \\
		{	EURO STOXX 50}&${\mathcal G}$+K& $3201.73$ &  3202.50&170.15&2918.05 &3485.50 \\

			\bottomrule
		\end{tabular}
\captionof{table}{\small Descriptive statistics.  Mean, median, standard deviation (std), and quantiles (q) 5\%, 95\% of some relevant quantities in the options' dataset we analyze. We report the number of strikes for each maturity and value date, the straddle position, $C(K)+ P(K)$, and the synthetic forward plus the strike,  ${\mathcal G}(K)+K$.}
\label{tab:stat descr}
\end{center}
The dataset also includes the OIS rates at 3:00 
p.m. London Time   (USD and EUR) with a time-to-maturity equal to 1-12, 15, 18, and 21 months and 2, 3, 4 and 5 years. 
The OIS interest rate curve is bootstrapped following the standard methodology \citep[see, e.g.][]{HenrardBook, baviera2015note}.
Eikon Reuters provides all financial data.  

The dataset provides call/put bid and ask prices for each available maturity. Data pre-processing criteria are simple:
we filter out the options that do not satisfy two basic liquidity criteria and we discard maturities with just one or two strikes. 
As first liquidity criterion, we filter the so-called ``penny options", i.e. options at a very low price. 
All options, whose value is less than 0.1  (S\&P 500 or EURO STOXX 50) index points, 
fall within this class. 
Then, options with a wide bid-ask spread are discarded. 
We filter out 
options with a ratio ask-bid/ask larger than 60\%.
This second liquidity criterion excludes strikes 
for which either bid or ask prices for call and put options are not available. 

\section{S\&P 500 and EURO STOXX 50 implicit interest rates}

In this section, we infer the market discount factor from option prices and analyze it for the two option markets. 
We verify whether the market discount factor corresponds to the EUR or 
USD OIS curve and find statistical evidence that a cost of funding of 34 basis points is added to the OIS curve in the S\&P 500 case. 
\\

In the market, we observe bid and ask prices for every different strike and a fixed maturity. 
The bid synthetic forward is obtained by selling the call and buying the corresponding put, vice-versa for the ask price.
Mid prices are the arithmetic average of bid and ask prices.
\begin{equation}
\label{forward_formula}
\left\{
 \begin{array}{rl}
     {\mathcal G}^{bid}\left(K\right)&:= \displaystyle {C^{bid}\left(K\right)-P^{ask}\left(K\right)}\\[4mm]
     {\mathcal G}^{ask}\left(K\right)&:= \displaystyle {C^{ask}\left(K\right)-P^{bid}\left(K\right)}\\[4mm]
      {\mathcal G}\left(K\right)\;\;\;\;\;&:=\displaystyle \frac{ {\mathcal G}^{bid}\left(K\right)+ {\mathcal G}^{ask}\left(K\right)}{2}   \; .
 \end{array}
\right.
\end{equation}

\begin{center}
\begin{minipage}[t]{1\textwidth}
\centering
{\includegraphics[width=.90\textwidth]{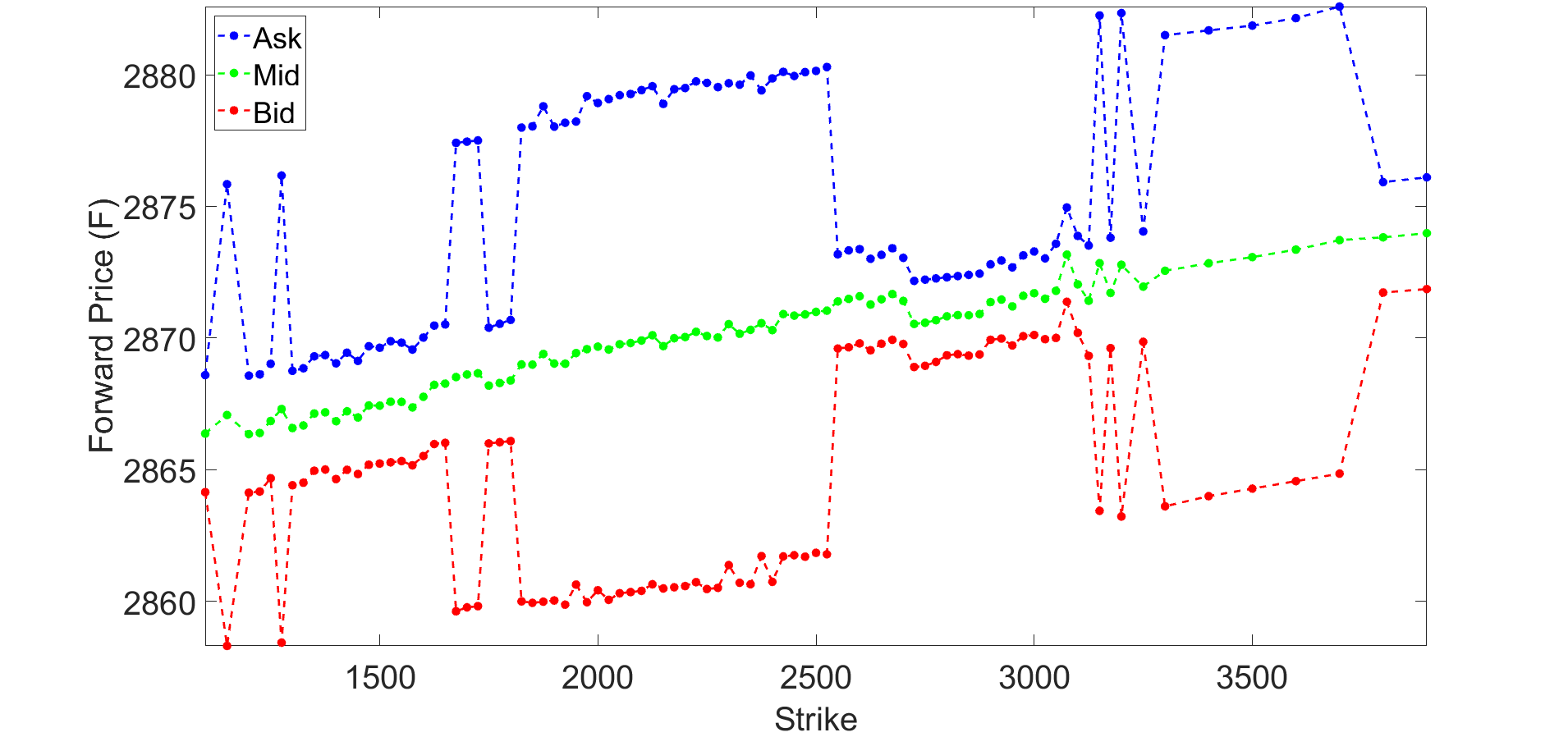}} 
\captionof{figure}{\small 
S\&P 500 
forward prices observed at 3:00 p.m. London Time of the $1^{st}$ of April 2019 with maturity on the $21^{st}$ of March 2020. 
Forward prices are obtained via synthetic forwards in (\ref{eq:Forward}) for different strikes: 
we assume $\overline{ B}(t_0,T) ={ B}(t_0,T) $, the  USD OIS discount factor.
We plot in red the bid prices, in blue the ask prices, and in green the mid prices. 
Notice that the mid prices are linear w.r.t. the strike. This fact denotes a market implied discount factor $\overline{ B}(t_0, T)$ lower than the USD OIS one, as explained in the text.
}
\label{fig:forwardOISIR}
\end{minipage}
\end{center}

In  \Cref{fig:forwardOISIR}, we plot an example of one year S\&P 500 forward prices $F$ in (\ref{eq:Forward}) using the discount factor ${ B}(t_0,T)$ of the USD OIS curve obtained 
via the bootstrap. 
We can notice a linear behavior w.r.t. the strikes.

 By non-arbitrage principle, the forward $F$ should be constant in $K$.
Thanks to equation (\ref{eq:Forward}), we observe from \Cref{fig:forwardOISIR} that also the ratio ${\mathcal G} {(K)}/{ B}(t_0,T)$ is a decreasing linear function  of $K$, but with an angular coefficient greater than $-1$, because it cannot compensate the linear term $K$ in (\ref{eq:Forward}).
Hence, 
in absolute value, the actual angular coefficient of ${ {\mathcal G} (K)}/{\overline{ B}(t_0,T)}$ should be larger that the one of ${\mathcal G} {(K)}/{ B}(t_0,T)$:
we infer that the actual discount $\overline{ B}(t_0,T)$ is lower than the OIS one ${ B}(t_0,T)$.

\bigskip

The discount factor used in the  market  $\overline{ B}(t_0,T)$ can be obtained as the angular coefficient in the linear regression 
\begin{equation}
{\mathcal G}_i = - \overline{ B}(t_0,T) \, K_i + \overline{ B}(t_0,T) \, F + \epsilon_i  \quad \quad  i=1,..,N
\end{equation}

for the different strikes $\{ K_i \}_{i=1,..,N}$ available at value date $t_0$ and maturity $T$, where $ \epsilon_i $ are some error variables.
Its least squares estimation is
\begin{equation}
\overline{ B}(t_0,T) = \displaystyle  -\frac{\sum^N_{i=1} (K_i - \hat{K}) ({\mathcal G}_i  -  \hat{{\mathcal G}} ) }{\sum^N_{i=1} (K_i - \hat{K})^2} \;
\end{equation}
where
\begin{equation}
	 \hat{{\mathcal G}} := \frac{1}{N} \sum^N_{i=1} {\mathcal G}_i  \quad , \quad \hat{K} := \frac{1}{N} \sum^N_{i=1} K_i \; .
\end{equation}
We observe that the regressions are very precise with an $R^2$ above $0.9995$
for all value dates $t_0$ and all maturities $T$ in the  dataset analyzed.

\bigskip

This result is equivalent to state that
a spread is added to the USD OIS curve. 
The funding spread (or cost of funding) can be defined in several ways; the simplest one is
\be
s := \frac{1}{T-t_0} \ln \frac{{ B}(t_0,T)}{\overline{ B}(t_0,T)} 
\label{eq:spread}
\en 
where ${ B}(t_0,T)$ has been obtained from the bootstrap of the OIS curve and time intervals are measured according to an 
$Act/365$ convention.\footnote{This is equivalent --up to a fraction of basis point-- to consider a cost of funding $s$ over the overnight rate OR.}
The  elementary indicator (\ref{eq:spread}) allows the market maker to monitor in real-time the cost of  funding in the derivative market where he operates; 
it allows also to detect possible situations of stress in funding liquidity.

\bigskip

We measure this spread for all value dates $t_0$ and all maturities $T$ in the whole options' dataset.
In  \Cref{fig:SpreadUS} we plot  the spread over the 
USD OIS curve w.r.t. the  synthetic  forward  time-to-maturity (ttm) $T-t_0$. 
It seems that a spread of 34 basis points is applied to the USD OIS curve for maturities higher than one month.

\begin{center}
\begin{minipage}[t]{1\textwidth}
\centering
{\includegraphics[width=.90\textwidth]{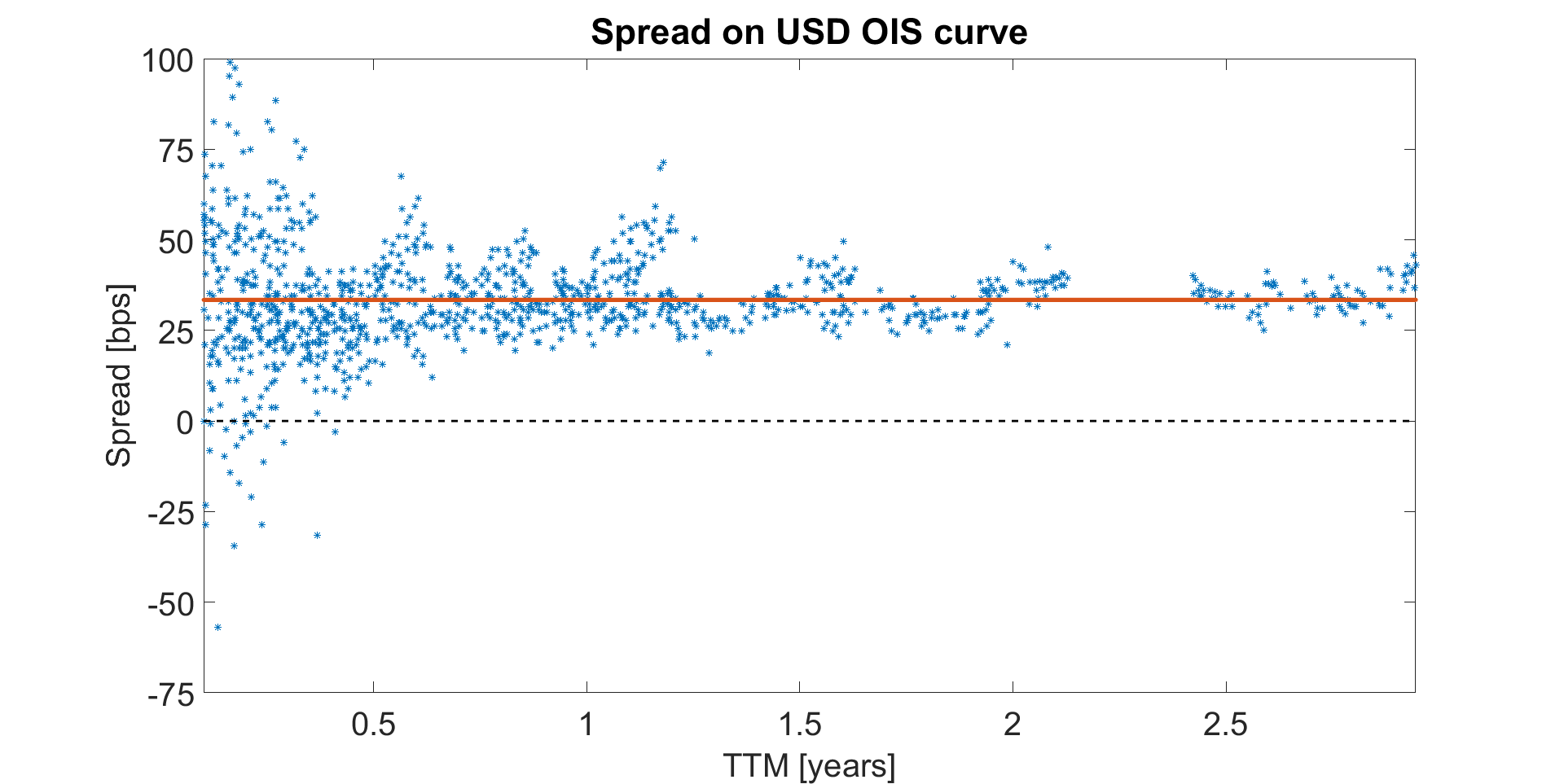}} 
\captionof{figure}{\small 
Spread over USD OIS in the S\&P 500 case. The spread (\ref{eq:spread}) is plotted  against the  time-to-maturity (ttm), 
for ttm longer than one month. 
The average spread of 34 basis points over the OIS curve seems constant over the different maturities (continuous red line).  
We observe a higher variance for short term maturities. 
}
\label{fig:SpreadUS}
\end{minipage}
\end{center}
We fit the spread $s$ as a function of the ttm and we test the statistical significance of the results.
We can accept the null hypothesis of no slope with a p-value of 11\% and we reject the null hypothesis of zero intercept with a p-value below $10^{-16}$. The intercept estimated assuming no slope is of 34 basis points.\\

We follow the same procedure for the EURO STOXX 50 forward prices, the spread over the EUR OIS curve is reported in \Cref{fig:SpreadEU}. 
\begin{center}
\begin{minipage}[t]{1\textwidth}
\centering
{\includegraphics[width=.90\textwidth]{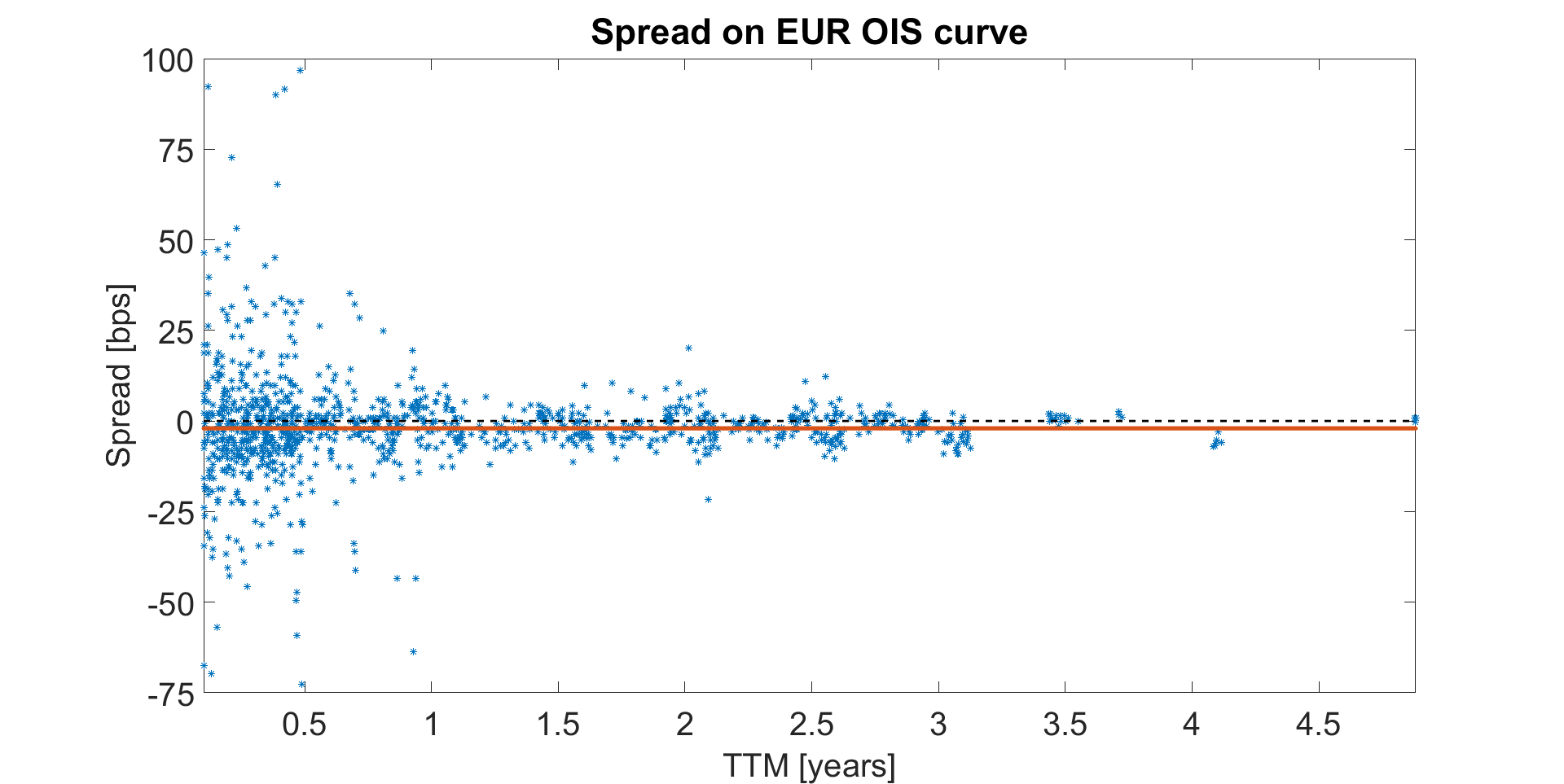}} 
\captionof{figure}{\small 
Spread over the EUR OIS curve in the EURO STOXX 50 case. The spread  (\ref{eq:spread}) is plotted against the ttm, 
for ttm larger than one month. The average spread  seems to be zero over the different maturities  (continuous red line).
}
\label{fig:SpreadEU}
\end{minipage}
\end{center}

We accept the null hypothesis of no intercept with a p-value of 23\% and we accept the null hypothesis of no slope with a p-value of 81\%. In  \Cref{tab:stat tests} we report a summary of the estimated slope and intercept parameters together with the statistical test p-values for both option markets.
We can conclude that dealers in the S\&P 500 market are subjected to a cost of funding, constant w.r.t. the ttm, on average   of 34 basis points, 
the same does not apply for dealers in the EURO STOXX 50 market.
 
\begin{center}
\begin{tabular} {cccccc}
			\toprule
			\textbf{market} & \textbf{parameter} &\textbf{estimate} & \textbf{p-value} \\ 
			\toprule
			{S\&P 500}&Intercept & 33& $<10^{-16}$ \\

			{S\&P 500} &Slope &  $1$ & $ 0.11$  \\
			\hline
			{EURO STOXX 50}&Intercept& $-1$ &  $      0.23$ \\
			
			{EURO STOXX 50}&Slope& $0$ & $   0.81$  \\
			
			\bottomrule
		\end{tabular}
\captionof{table}{\small Spread over OIS. Estimated intercept and slope of the spread over the OIS curve in basis points 
(USD OIS curve for S\&P 500 and EUR OIS curve for EURO STOXX 50). We accept the null hypotheses of no slope for both markets. We refuse the null hypothesis of zero intercept only for the S\&P 500 market. There is statistical evidence that dealers in the S\&P 500 are subjected to a cost of funding, constant w.r.t. the ttm: the intercept estimated assuming no slope is of 34 basis points. No spread is observed for the   EURO STOXX 50.}
\label{tab:stat tests}
\end{center}

\smallskip

We observe in both Figure \ref{fig:SpreadUS} and \ref{fig:SpreadEU} a higher variance for short maturities. 
This is due to the fact that only the product of the spread and the time-to-maturity is relevant for the forward: 
for shorter maturities, the no-arbitrage condition is granted by a larger range of values for the spread. \\

Four robustness tests are performed. (i) We fit a weighted linear regression \citep[see, e.g.][Ch.3, p.51]{strutz2010data}  to tackle heteroskedasticity problems.
 The weights are selected as one over the square of the  linear regression residuals.
(ii) We change the penny-option and the bid-ask spread thresholds respectively in the range (0, 1) and (30\%, 90\%) to verify the robustness w.r.t. the excluded strikes.
(iii) We extend the analysis window  up to the $1^{st}$ of October 2019 
(the last date before the EONIA is discontinued\footnotemark[2]) and limit the analysis to ttm larger than either 6 or 12 months.
(iv) We do not discard the maturities with less than three valid strikes.
  In all robustness tests, results do not change up to a single basis point.

\bigskip

Let us underline that this methodology allows us to determine also the forward price obtained via synthetic forwards.
This forward price is obtained from the put-call parity relation (\ref{eq:Forward}) using the $\overline{B}(t_0,T)$ that includes the cost of funding:
at a given maturity $T$, the forward ask price is 
the lowest forward ask in (\ref{eq:Forward}) and the forward bid price is the highest forward bid. 
In \Cref{fig:forward curve} we show an example of the forward bid and ask prices obtained in this way.
\begin{center}
\begin{minipage}[t]{1\textwidth}
\centering
{\includegraphics[width=.90\textwidth]{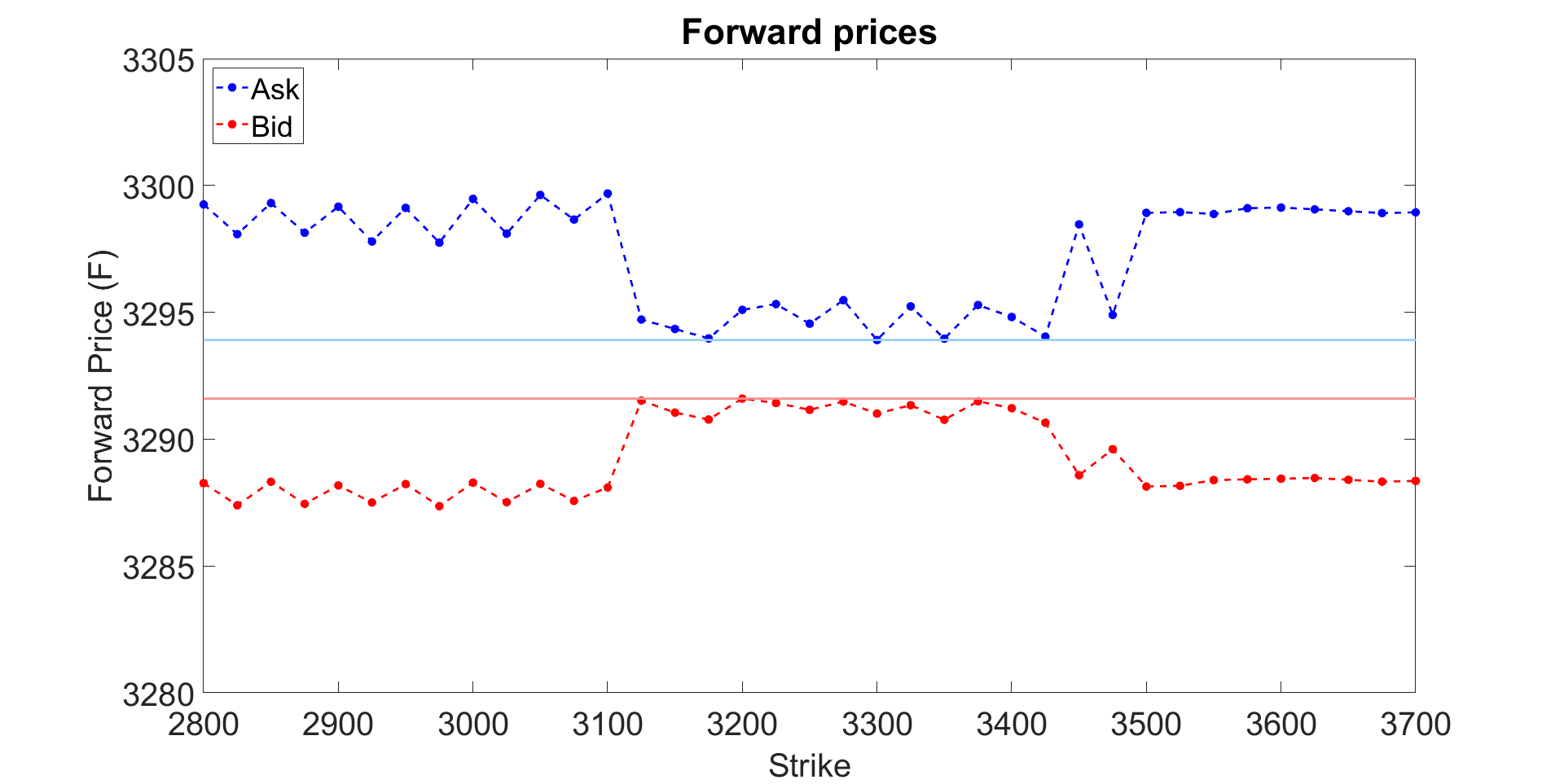}} 
\captionof{figure}{\small 
 Example of the construction of the forward price via synthetic forwards. We show
bid (in red) and ask (in blue) forward prices (\ref{eq:Forward}) of the EURO STOXX 50 at the $1^{st}$ of April 2019 for the 2$1^{st}$ of June 2019 maturity. 
Only prices not discarded by the two liquidity criteria described in the text are considered. 
We also plot the forward ask price (continuous light blue) and bid price (continuous light red) obtained as the lowest and highest values respectively.   Notice that the length of the bid-ask interval changes with the strikes signaling different liquidity for different strikes. 
}
\label{fig:forward curve}
\end{minipage}
\end{center}

\section{Conclusions}


Which discount factor should be used in exchange-traded derivatives? This study exploits the implications of the put-call parity to develop a methodology that allows us to recover the discount factor implied by option prices on S\&P 500 and EURO STOXX 50 indices. 
A dealer in the option market can use this technique to real-time monitor the funding spreads of market players. 
The implicit discount factor is the one such that a
forward contract, built using the put-call parity relation,  does not depend on the strike. 
We compute the S\&P 500 and EURO STOXX 50 option markets' implicit discount factors and evaluate the cost of funding over the 
curve obtained bootstrapping OIS derivatives.
We have statistical evidence of  a cost of funding of, on average,  34 basis points  on top of the 
USD OIS curve in the S\&P 500 case and no cost of funding for the EURO STOXX 50 case. 
This cost of funding is constant for all liquid maturities up to several years for both markets. \\

Hence, the natural question is: why do we observe a spread over USD and no spread over EUR OIS?
The reason should be sought in the differences between the two underlying money markets.
Let us remind that the FED Target range indicates only some target rates, 
while in Europe the corridor system denotes the real rates at which ECB serves as lender of last resort to the financial system. 
In the USD market the two rates, the collateralized one (SOFR) and the uncollateralized one (EFFR), 
differ for a spread that can be significant in several days. A dysfunctional repo market, 
indicated by a sharp spike in the SOFR, has been observed several days in the analyzed 
period;\footnote{For example, 
during the period of analysis starting from the $1^{st}$ of November 2018 and ending on the 
$1^{st}$ of October 2018, apart from the end of months (and, in particular, the End-of-Year),
spikes outside the FED Fund range are observed in $15$ days. Let us notice that, in all these days, SOFR is always larger than the upper side of the range.} 
besides, a disruption in the repo market has been signalled by the well-known repo blow-up of the $17^{th}$ of September 2019, 
when the spread between the two fixings (SOFR and EFFR) topped to almost 3 percent and 
prompted the Federal Reserve to inject tens of billions of dollars of reserves into money markets \citep[see e.g.][]{Barret, Tilford}.
While, as we have already underlined in the introduction, the OIS market is mainly based on the EFFR, 
the volumes in the money market are mostly concentrated on the repo rate, with SOFR volumes more than ten times larger than EFFR ones 
\citep[see, e.g.][]{BIS_Liquidity1}.
These funding disruptions in the USD money market, not observed in the EUR market, suggest that 
market players could require a spread over OIS as a premium for the additional liquidity risk observed in this market.
As for future research, one main promising direction is evident:
it could be interesting to understand whether this funding spread is connected to the implied/historical volatility on the two indices.


\section{Acknowledgments}
The authors are grateful to three anonymous referees for their helpful feedback and suggestions.
We also thank M. Beretta (Gardena Capital), A. Cassaro and T. Santagostino Baldi (Goldman Sachs), L. Fusco (Unicredit), F. Gregori (BNPP), E. Mercuri (Banor) and M. Spadaccino (Illimity) for enlightening discussions on the topic.
\bibliography{sources}
\bibliographystyle{tandfx}
\end{document}